\documentclass[journal,comsoc,draftclsnofoot,onecolumn,12pt]{IEEEtran}

\usepackage{amsmath}
\usepackage{amssymb}
\usepackage[pdftex]{graphicx}
\usepackage{graphics}
\usepackage[T1]{fontenc}
\usepackage{cite}
\usepackage[cmintegrals]{newtxmath}
\usepackage{array}
\usepackage{bbm}
\usepackage{multirow}
\usepackage{makecell}
\usepackage{xcolor,colortbl}

\DeclareMathAlphabet{\mathcal}{OMS}{cmsy}{m}{n}

\DeclareMathOperator{\RE}{Re}
\DeclareMathOperator{\IM}{Im}

\newcommand{\comment}[1]  {}
\def\BE{\begin{equation}}
\def\EE{\end{equation}}
\def\BEA{\begin{eqnarray}}
\def\EEA{\end{eqnarray}}
\newcommand{\pd}[2]{\frac{\partial #1}{\partial #2}}

\newcommand\vh{{\bf h}}

\newcommand\vm{{\bf m}}
\newcommand\vn{{\bf n}}
\newcommand\vo{{\bf o}}

\newcommand\vr{{\bf r}}
\newcommand\vs{{\bf s}}

\newcommand\vv{{\bf v}}

\newcommand\vx{{\bf x}}
\newcommand\vy{{\bf y}}
\newcommand\vz{{\bf z}}

\newcommand\mA{{\bf A}} 

\newcommand\mH{{\bf H}}
\newcommand\mI{{\bf I}}

\newcommand\mM{{\bf M}}

\newcommand\mR{{\bf R}}

\newcommand\ie{{{i.e.}}}
\newcommand\eg{{{e.g.}}}
\DeclareMathOperator*{\argmin}{argmin}
\definecolor{Gray}{gray}{0.85}
\newcolumntype{a}{>{\columncolor{Gray}}c}
\newcommand{\snr}{ {\mathsf{snr}} }

\begin{document}
%
\title{Massive BLAST: An Architecture for Realizing Ultra-High Data Rates for Large-Scale MIMO}
%
%
%

\author{Ori~Shental,~\IEEEmembership{Member,~IEEE,}
        Sivarama Venkatesan,~\IEEEmembership{Member,~IEEE,}
        Alexei~Ashikhmin,~\IEEEmembership{Fellow,~IEEE,}
        and~Reinaldo~A.~Valenzuela,~\IEEEmembership{Fellow,~IEEE}
\thanks{O. Shental, S. Venkatesan, A. Ashikhmin and R.A. Valenzuela are with Nokia Bell Laboratories, NJ, USA, e-mail: \{ori.shental,venkat.venkatesan,alexei.ashikhmin,reinaldo.valenzuela\}@nokia-bell-labs.com.}
}


\maketitle

\begin{abstract}
A detection scheme for uplink massive MIMO, dubbed massive-BLAST or M-BLAST, is proposed. The derived algorithm is an enhancement of the well-known soft parallel interference cancellation. Using computer simulations in massive MIMO application scenarios, M-BLAST is shown to yield a substantially better error performance with reduced complexity, compared to the benchmark alternative of a one-shot linear detector, as well as the original sequential V-BLAST. Hence, M-BLAST may serve as a computationally efficient means to exploit the large number of antennas in massive MIMO.
\end{abstract}


\IEEEpeerreviewmaketitle

\section{Introduction}
%
%
%
%
\IEEEPARstart{T}{he} introduction of the V-BLAST (Vertical-Bell Laboratories Layered Space-Time,~\cite{wolniansky1998v}) detection algorithm was one of the main enablers of the vast proliferation of multiple-input multiple-output (MIMO) systems over the last two decades. Massive MIMO, also known as large-scale MIMO~\cite{marzetta2010noncooperative,marzetta2016fundamentals}, is a scalable version of point-to-point MIMO, or multiuser MIMO, with many antennas at both link ends.

The current detection paradigm in massive MIMO mainly relies on (one-shot) linear signal processing schemes such as the matched-filtering, zero-forcing (ZF) and minimum-mean-square-error (MMSE) detectors. As a manifestation of successive interference cancellation (SIC), V-BLAST is not practically extendable to massive MIMO systems since the number of iterations required to peel off the various layers increases with the number of transmitting antennas.

As a potential remedy one may consider the utilization of parallel interference cancellation (PIC, or multistage detector~\cite{varanasi1990multistage}). A soft hyperbolic tangent decision version of PIC is known to be asymptotically optimal in the large-system limit~\cite{tanaka2005approximate}, assuming it converges. However the latter suffers from relatively slow convergence rate especially in a realistic signal-to-noise-ratio (SNR) regime of operation.

In this letter we propose a detection algorithm for large-scale MIMO, dubbed massive-BLAST (or M-BLAST). Our derivation of M-BLAST relies on a more accurate approximation of the logarithm of the partition function appearing in the underlying Bayesian inference problem. Such an approximation has deep roots in the statistical mechanics literature (see, \eg,~\cite{yedidia2001idiosyncratic} for a comprehensive overview). Furthermore, the derived M-BLAST can be conceived as an improvement of the conventional PIC, yielding a substantially more accurate inference of the large number of concurrently transmitted bits.
Based on simulations of a massive MIMO uplink channel, M-BLAST is shown to yield significantly better error performance. Consequently, higher throughput is demonstrated across the entire simulated SNR range, compared to not only the benchmark one-shot linear detectors but also the seminal V-BLAST.

The letter is organized as follows. The massive MIMO system model is described in Section~\ref{sec_model}. Section~\ref{sec_MBLAST} derives the M-BLAST algorithm and Section \ref{sec_results} discusses the simulation results for the error performance and throughput gains. Finally, Section~\ref{sec_conclusion} contains some concluding remarks.

We shall use the following notations. The superscript $T$
denotes a vector or matrix transpose, $\mI_{D}$ denotes a $D\times D$ identity matrix, and the symbols $M_{ij}$ and $v_{i}$ denote the $ij$th and $i$th scalar entries of the matrix $\mM$ and vector $\vv$, respectively. The operator $\mathbbm{E}(\cdot)$ refers to expectation w.r.t.~the distribution of the input argument, while $\text{diag}(\mM)$ is a matrix populated with the diagonal entries of $\mM$. The symbols $\RE(\cdot)$ and $\IM(\cdot)$ denote the real and imaginary parts of a complex argument, respectively. The operation $\langle\vv\rangle$ means averaging over the entries of the vector $\vv$.

\section{System Model}\label{sec_model}

Consider a basic multiuser MIMO~\cite{marzetta2016fundamentals} uplink channel with $K$ transmitting users (single antenna each) and $M$ receiving antennas at the base station. The MIMO channel adheres to
\BE\label{eq_model}
    \vr = \tilde{\mH}\mA\vs+\vv,
\EE
where $\vs\in\{\pm1,\pm j\}^{K}$ is a $\pi/4$-QPSK input vector. The complex fading channel matrix $\tilde{\mH}\in\mathbbm{C}^{M\times K}$ has i.i.d. entries with zero mean and variance $\frac{1}{M}$, to keep the received SNR independent of the number of receiving antennas. The vector \mbox{$\vv\sim\mathcal{CN}(\mathbf{0},\sigma^2\mI_{M})$} is a complex Gaussian noise vector. Let the root-power matrix $\mA\in\mathbbm{R}^{K\times K}$ be a diagonal matrix with $A_{ii}=\sqrt{P_{i}}$, where $P_i$ encapsulates the transmit power of the $i$th user and path loss, including large-scale fading. Finally, $\vr\in\mathbbm{C}^{M}$ is the received vector. The matrices $\tilde{\mH}$ and $\mA$, and noise variance $\sigma^{2}$ are assumed to be known, either perfectly or approximately, by the base station.
As is typical of a massive MIMO uplink, the number of concurrently transmitting antennas, $K$, is much smaller than the number of receiving antennas, $M$. For derivation purposes, hereinafter we assume a large-system limit $K,M\to\infty$ with a fixed ratio $\beta\triangleq\frac{K}{M}\in\mathbbm{R}<1$.


To ease the M-BLAST derivation, we shall replace the complex-valued model~\eqref{eq_model} by its real-valued equivalent
\BE\label{eq_model2}
\vy=\mH\vx+\vn.
\EE
To this end, we define the $2K$-dimensional vector $\vx$, and the $2M$-dimensional vectors $\vy$ and $\vn$, to be composed of a concatenation of the real and imaginary parts of $\vs$, $\vr$, and $\vv$, respectively.
We also define the real $2M\times2K$ matrix
\BE
\mH\triangleq\left(
               \begin{array}{cc}
                 \RE{(\tilde{\mH})} & \IM{(\tilde{\mH})} \\
                 -\IM{(\tilde{\mH})} & \RE{(\tilde{\mH})} \\
               \end{array}
             \right)\left(
                      \begin{array}{c}
                        \mA \\
                        \mA \\
                      \end{array}
                    \right)
             .
\EE

\section{M-BLAST Derivation}\label{sec_MBLAST}
The posterior probability associated with the channel model~\eqref{eq_model2} can be written as
\BE
\Pr(\vx|\vy,\mH,\sigma^2)=\frac{1}{\mathcal{Z}}\exp\Big(\sum_{i=1}^{j-1}\sum_{j=1}^{2K}R_{ij}x_{i}x_{j}+\sum_{i=1}^{2K}h_{i}x_{i}\Big),
\EE
where $\mR\triangleq-\frac{2\mH^{T}\mH}{\sigma^2}$ and $\vh\triangleq\frac{2\mH^{T}\vy}{\sigma^2}$. The \emph{normalizing partition function} is defined as (see, for instance,~\cite[Sections 1.2-1.3]{yedidia2001idiosyncratic})
\BE
\mathcal{Z}\triangleq\sum_{\vx\in\{\pm1\}^{2K}}\exp\Big(\sum_{i=1}^{j-1}\sum_{j=1}^{2K}R_{ij}x_{i}x_{j}+\sum_{i=1}^{2K}h_{i}x_{i}\Big).
\EE

Denoting $\mathcal{F}(\hat{\vm})\triangleq-\ln(\mathcal{Z})$ (the negative logarithm of the partition function, also known as \emph{free energy}), the desired vector of marginal posterior expectations is given, in the large-system limit, by~\cite[Sections 1.4-1.5]{yedidia2001idiosyncratic})
\BE\label{eq_min}
    \hat{\vm}\triangleq\mathbbm{E}(\vx|\vy,\mH,\sigma^2)=\argmin_{\vm}\mathcal{F}(\vm).
\EE
Here $\vm$ is the vector of expectations w.r.t.~some arbitrary $\tilde{\Pr}(\vx|\vy,\mH,\sigma^2)$, which is a tractable distribution approximating the actual intractable posterior distribution $\Pr(\vx|\vy,\mH,\sigma^2)$, with corresponding expectation vector $\hat{\vm}$.
Following the procedure described in~\cite[Section 2]{plefka1982convergence}, we replace $\mR$ by $\lambda\mR$ in the free energy expression (later we will set the auxiliary scalar $\lambda=1$). Leaving, for now, convergence issues aside, the corresponding partition function's logarithm is approximated via a Taylor expansion w.r.t.~$\lambda$ as
\BE\label{eq_F_Taylor}
    \mathcal{F}(\vm,\lambda)=\mathcal{F}_{0}(\vm)+\lambda\mathcal{F}_{1}(\vm)+\frac{\lambda^2}{2!}\mathcal{F}_{2}(\vm)+\ldots,
\EE
with $\mathcal{F}_{n}(\vm)\triangleq\frac{\partial^{n}}{\partial\lambda^{n}}\mathcal{F}(\vm,\lambda)\Big|_{\lambda=0}$.
This yields
\BEA
    \mathcal{F}_{0}(\vm)&=&\sum_{i=1}^{2K}\Big\{\frac{1+m_{i}}{2}\ln\frac{1+m_{i}}{2}+\frac{1-m_{i}}{2}\ln\frac{1-m_{i}}{2}\Big\},\nonumber\\\\
    \mathcal{F}_{1}(\vm)&=&-\sum_{i=1}^{j-1}\sum_{j=1}^{2K}R_{ij}m_{i}m_{j}-\sum_{i=1}^{2K}h_{i}m_{i},\\
    \mathcal{F}_{2}(\vm)&=&-\frac{1}{2}\sum_{i=1}^{j-1}\sum_{j=1}^{2K}R_{ij}^{2}(1-m_{i}^{2})(1-m_{j}^2).\label{eq_Onsager}
\EEA

Hence, according to~\eqref{eq_min}, minimizing the free energy approximation~\eqref{eq_F_Taylor} w.r.t.~$\vm$ for $\lambda=1$, one gets for any $i=1,\ldots,2K$ (following similar steps as in~\cite[Sections 1.5-1.6]{yedidia2001idiosyncratic})
\BEA\label{eq_derivative}
   0=\pd{\mathcal{F}(m_{i},\lambda=1)}{m_{i}}&=&\tanh^{-1}(m_{i})-\sum_{i\neq j=1}^{2K}R_{ij}m_{j}-h_{i}\nonumber\\&+&
   \sum_{i\neq j=1}^{2K}R_{ij}^{2}(1-m_{j}^2)m_{i}.
\EEA
In the last term of \eqref{eq_derivative}, $R_{ij}^{2}$ can be approximated by its average which is exact in the large-system limit. Now, explicitly expressing the solution to~\eqref{eq_derivative} in terms of the desired vector of marginal expectations, $\vm$, yields the following self-consistency equations for an iterative index $t\in\mathbbm{Z}_{\ge0}$
\BEA
    \vm^{t+1}&=&\tanh\Big(\frac{2}{\sigma^{2}}(\mR_{\smallsetminus}\vm^{t}+\mH^{T}\vz^{t})\Big),\label{eq_m}\\
    \vz^{t}&=&\vy-\mH\vm^{t}+\vo^{t-1},\label{eq_z}\\
    \vo^{t-1}&\triangleq&\beta\vz^{t-1}\Big(1-\Big\langle(\vm^{t})^{2}\Big\rangle\Big)\label{eq_o},
\EEA
with the diagonal matrix $\mR_{\smallsetminus}\triangleq\text{diag}\{\sigma^{2}\mR/2\}$. The initial conditions for $t\in\mathbbm{Z}_{<0}$ are $\vm^{t+1}=\vz^{t}=\vo^{t-1}=\mathbf{0}$. From these fixed-point equations an approximation to the desired posterior expectation can be inferred after a predetermined number of iterations.

Note that arbitrarily setting $\vo^{t-1}=\mathbf{0}$ (\ie, removing~\eqref{eq_o} from the set of equations), the fixed-point equations~\eqref{eq_m}-\eqref{eq_z} boil down to the well-known soft PIC.  Therefore in this sense, for a given number of iterations, the derived M-BLAST scheme can be viewed simply as an improvement of the conventional PIC. Furthermore, complexity-wise the computation of the additional term in M-BLAST, $\vo^{t-1}$~\eqref{eq_o}, requires only a straightforward and simple processing of already obtained information from previous iterations. This important addition originates from what is known as the Onsager correction term~\eqref{eq_Onsager}~\cite{yedidia2001idiosyncratic}.
Also note that convergence is guaranteed as long as \mbox{$\beta\Big(1-\Big\langle(\vm^{t})^{2}\Big\rangle\Big)<1$}.
Hence the proposed scheme is typically suitable for uplink underloaded massive MIMO scenarios with $\beta<1$.
Finally, note that the obtained iterative equations~\eqref{eq_m}-\eqref{eq_o} may provide a rationalization to recent literature on \emph{damped} interference cancellation schemes (\eg,~\cite{som2010improved}). Such schemes were originally established mainly on heuristics, thus heavily reliant on simulation-based optimization of the damping factor, rather than firm theoretical justification.

\section{Simulation Results and Discussion}\label{sec_results}
The proposed M-BLAST scheme \eqref{eq_m}-\eqref{eq_o} with $t=10$ iterations is simulated in an uplink Rayleigh flat-fading massive MIMO channel, $\tilde{H}_{ij}\sim\mathcal{CN}(0,\frac{1}{M})$, with $K=500$ BPSK transmitting users and $M=1000$ receiving antennas at the base station (thus the load $\beta=0.5$). The error performance of the M-BLAST, in bit-error-rate (BER), is compared to the non-fading single-input single-output (SISO) AWGN lower bound and to several conventional detectors: a one-shot linear MMSE, MMSE-based SIC (V-BLAST) and ordinary soft PIC, also with $t=10$ iterations. We first assume uncoded streams and users transmitting with equal SNRs (\ie, $\mA=\mI_{K}$).

\begin{figure}[!tb]
  \centering
    \includegraphics[width=0.7\columnwidth]{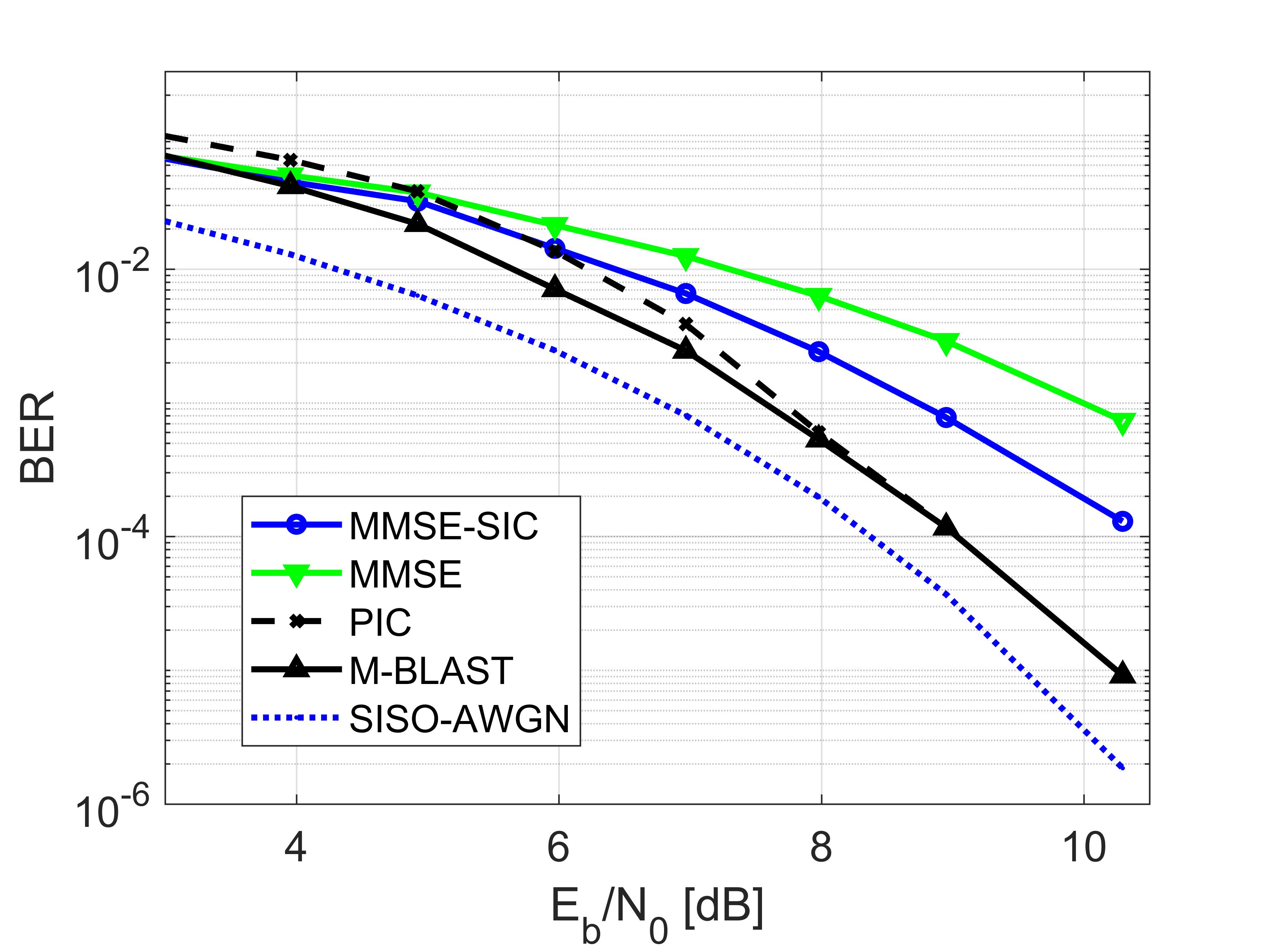}\vspace{-0cm}
  \caption{BER as a function of $E_{\text{b}}/N_{0}$ for $K=500$ equal SNR users with uncoded streams and imperfect CSI.}\label{fig_1}
\end{figure}

Fig.~\ref{fig_1} plots the BER versus $E_{\text{b}}/N_{0}$ for the different detectors, assuming imperfect channel state information (CSI) and non-ideal noise variance estimation at the base station. Imperfect CSI is modeled via $\hat{H}_{ij} = H_{ij}+\mathcal{N}(0,\frac{1}{2\snr_{p}})$, where $\snr_{p}$ is the pilot-symbol SNR (in LTE $\snr_{p}$ can be up to 6 dB above the data-symbol SNR). In order to model the fact that the base station has only approximate estimation of the noise variance, its estimate $\hat{\sigma}^{2}$ is randomly taken from a uniform distribution within the range $(1\pm X)\sigma^{2}$. In Fig.~\ref{fig_1}, $X=1\%$ is being used which is a typical value for static users in LTE.
M-BLAST is observed to yield a reduced BER across the entire examined $E_{\text{b}}/N_{0}$ range compared to the common MMSE and V-BLAST detectors. For the lower SNR levels M-BLAST exhibits non-negligible gain also over the soft PIC.
Quantitatively, for the operating point of $1\%$ uncoded BER, M-BLAST yields gains of approximately 2 dB over MMSE, 1 dB over V-BLAST, 0.7 dB over PIC, and is about 1 dB away from the SISO-AWGN bound. \comment{For a more moderate setup of $K=32$ users and $M=96$ receiving antennas, similar trend and gain in error performance of M-BLAST over the linear MMSE were observed (those results are not shown here due to lack of space).}

\begin{figure}[!tb]
  \centering
  \includegraphics[width=0.7\columnwidth]{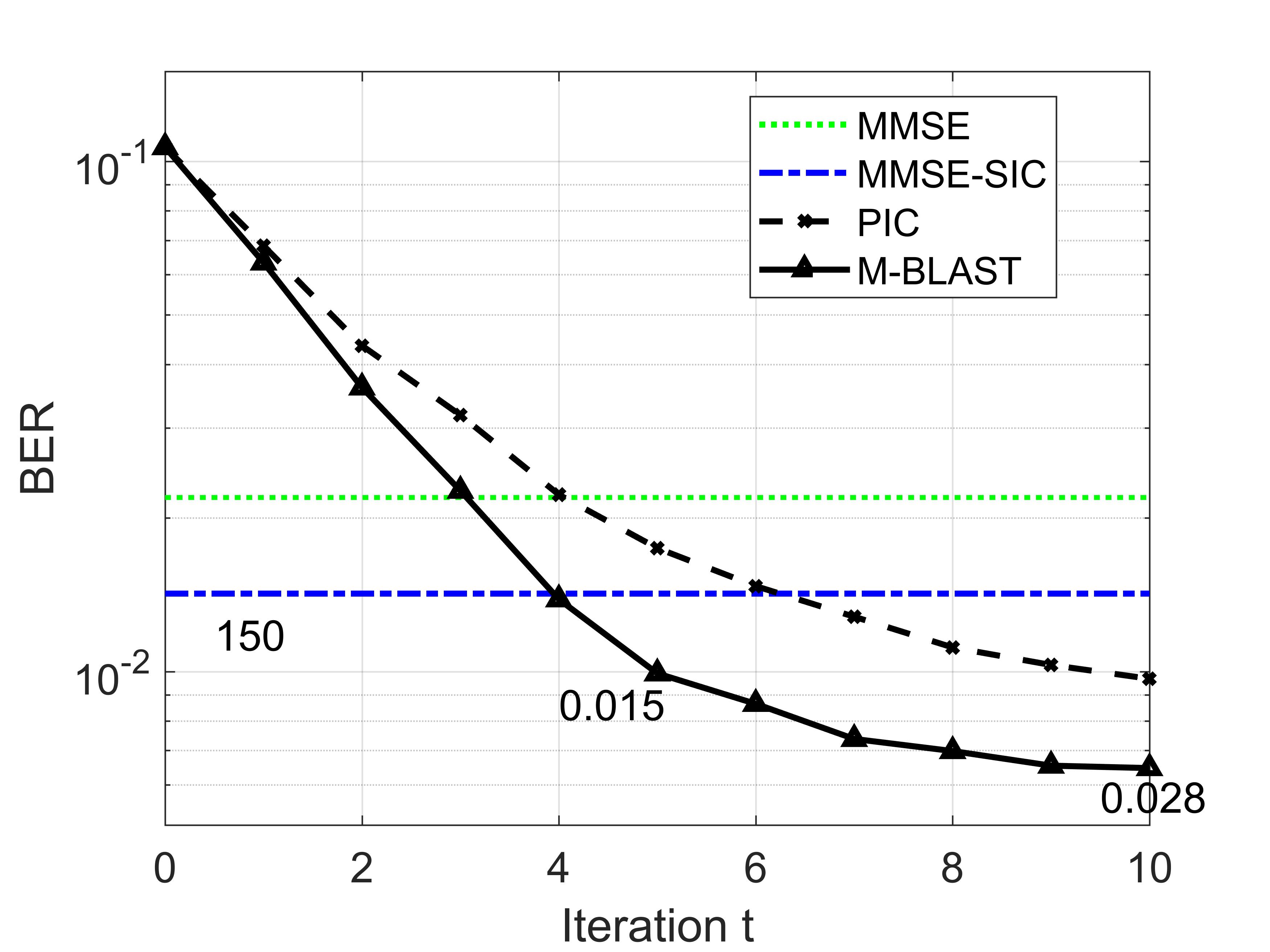}\vspace{-0cm}
  \caption{BER as a function of iterations $t$ at $E_{\text{b}}/N_{0}=6$ dB ($K=500$, $M=1000$ and imperfect CSI). The relative complexity w.r.t. MMSE is also marked for the V-BLAST and M-BLAST (for 5,10 iterations).}\label{fig_complex}
\end{figure}

\comment{Below we discuss the attractiveness of M-BLAST in terms of `multiply \& accumulate' (MAC) operations.} The complexity of the linear MMSE detector is primarily determined by the complexity of computing the Gram matrix, $\mH^{T}\mH$, and the matrix inversion. For practical algorithms these exact computations on non-structured dense matrices are typically cubic. Ignoring linear terms, the total complexity, in `multiply \& accumulate' (MAC) operations, explicitly scales as $O_{\text{MMSE}}(K,M)\triangleq\mathcal{O}(K^2M)+\mathcal{O}(K^3)+\mathcal{O}(KM)+\mathcal{O}(K^2)$. The last two non-leading order terms emerge from the matched-filtering, $\mH^{T}\vy$, and MMSE filtering, respectively. Evidently, in the era of millimeter-wave wireless communications with massive MIMO, the number of users, $K$, can be in the thousands, severely inflating the linear detector's complexity. For V-BLAST the computational burden is even worse as it inherently requires $K$ stages in order to infer the users' data. Hence it is governed by $\sum_{k=1}^{K}O_{\text{MMSE}}(K-k+1,M)\approx\mathcal{O}(K^4)$, that is, by the complexity of $K$ consecutive linear MMSE operations of diminishing sizes.
However, the number of MAC operations in the M-BLAST architecture is mainly dominated by matrix-vector multiplications ($\mH^{T}\vz^{t}$ in~\eqref{eq_m} and $\mH\vm^{t}$ in~\eqref{eq_z}) of quadratic complexity and is only linear in the number of iterations $t$. Thus, again ignoring linear terms, the complexity of M-BLAST scales as $O_{\text{M-BLAST}}(K,M)\triangleq\mathcal{O}(2tKM)$. \comment{Also note, based on the additional term~\eqref{eq_o}, that M-BLAST is only marginally more complex than the conventional soft PIC. However, as illustrated on Fig.~\ref{fig_complex}, this term facilitates the faster convergence and improved error performance when compared to PIC.}

Fig.~\ref{fig_complex} compares the uncoded BER of M-BLAST \eqref{eq_m}-\eqref{eq_o} and PIC (only~\eqref{eq_m}-\eqref{eq_z}) as a function of the number of iterations $t$ for a setup similar to the one used in Fig.~\ref{fig_1} at a particular $E_{\text{b}}/N_{0}=6$ dB. The faster convergence of M-BLAST over PIC, driven by the additional term~\eqref{eq_o}, is apparent, where in this case $t=10$ seems to be sufficient for M-BLAST to converge. Also drawn are the BER of MMSE and MMSE-SIC. One can see that M-BLAST beats the two detectors after only 3 and 4 stages, respectively. A straightforward enumeration of MAC operations shows that in such a large-scale setup ($K=500$, $M=1000$) V-BLAST is 150 times more complex than the one-shot MMSE, while M-BLAST costs less than $2\%$ (resp.~$3\%$) of the MAC operations of MMSE for $t=5$ (resp.~$t=10$) iterations. \comment{A similar enumeration for a more moderate setup of $K=32$ and $M=96$ shows that MMSE-based V-BLAST requires 10 times more MAC operations than the one-shot linear MMSE, while the proposed M-BLAST with $t=5$ (resp.~$t=10$) iterations requires only 1/4 (resp.~1/2) the operations.}
Table~\ref{tab_2} summarizes the simulated 1\% uncoded BER gains and the corresponding complexity reductions of M-BLAST over linear MMSE for different, modest to very large, uplink massive MIMO configurations. For a nowadays practical configuration of $K=8$ and $M=64$ the complexity of the two schemes is comparable, while M-BLAST exhibits about 0.4 dB gain. For very large MIMO systems M-BLAST delivers not only an impressive complexity advantage, but also substantial gains.

\begin{table}[!tb]
    \renewcommand{\arraystretch}{1}
\caption{\scriptsize{M-BLAST gains over MMSE in 1\% BER and complexity (imperfect CSI).}}\vspace{-0cm}
\centering
\resizebox{0.55\columnwidth}{!}{%
\tiny
\begin{tabular}{|c||c|c|c|c|c|}
  \hline
  Users $K$ & 8 & 32 & 64 & 500 & 1000 \\\hline
  Rx. Antennas $M$ & 64 & 96 & 192 & 1000 & 2000 \\\hline
  Load $\beta$ & 1/8 & 1/3 & 1/3 & 1/2 & 1/2 \\\hline
  M-BLAST iterations $t$ & 5 & 5 & 5 & 10 & 10 \\\hline
  $\frac{O_{\text{M-BLAST}}}{O_{\text{MMSE}}}$ [\%] & \textbf{99} & \textbf{23} & \textbf{12} & \textbf{2.7} & \textbf{1.3} \\\hline
  {\makecell{1\% BER\\Gain [dB]}} & \textbf{0.4} & \textbf{0.9} & \textbf{1} & \textbf{1.8} & \textbf{2} \\
  \hline
\end{tabular}\label{tab_2}
}
\end{table}

\begin{figure}[!tb]
  \centering
  \includegraphics[width=0.7\columnwidth]{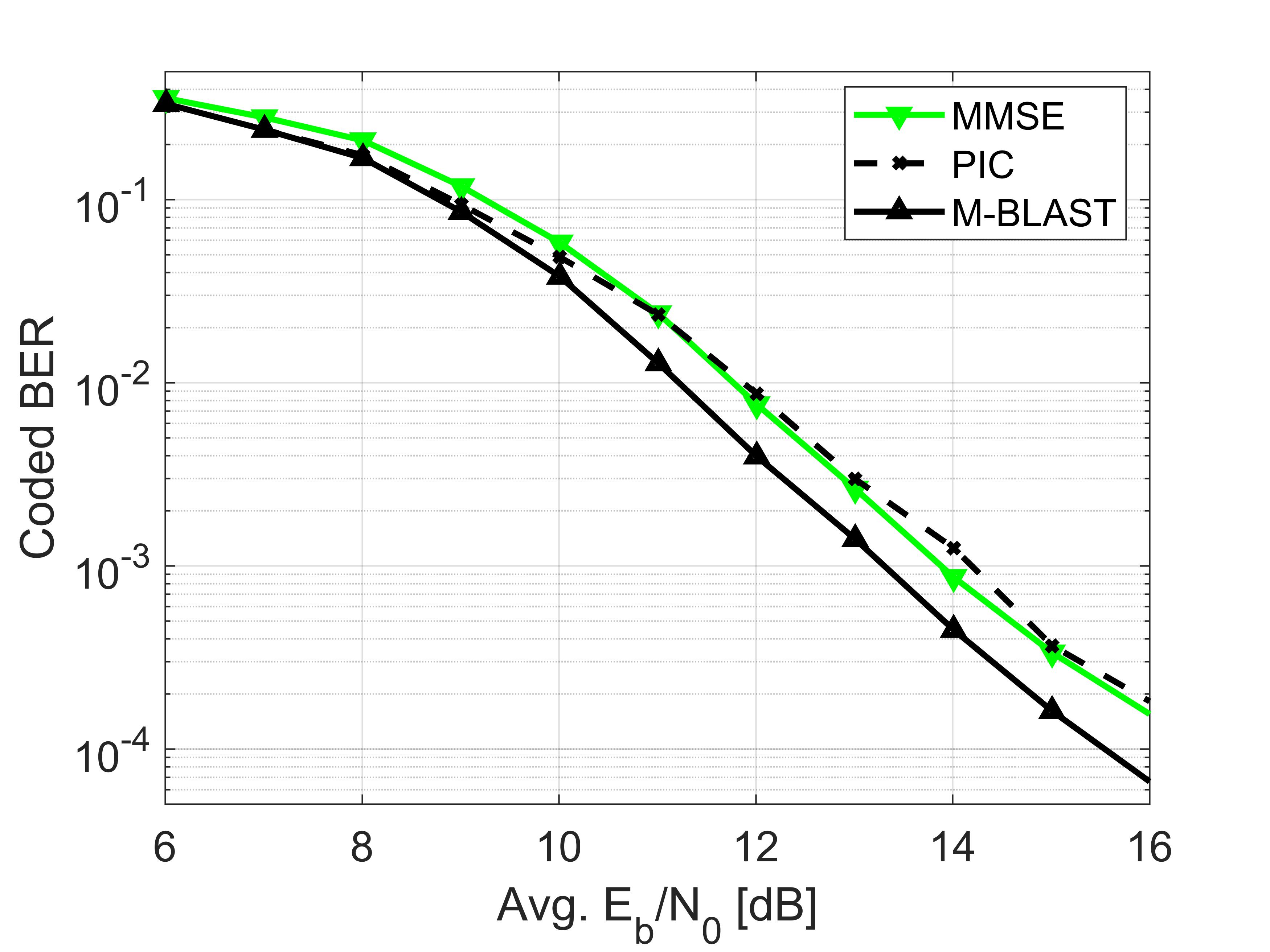}\vspace{-0cm}
  \caption{Coded BER as a function of average $E_{\text{b}}/N_{0}$ for $K=500$ non-equal SNR users with coded streams and perfect CSI.}\label{fig_3}
\end{figure}

We now consider the case of users with unequal SNRs. For this purpose the users' SNRs are randomly generated from a cumulative distribution function (CDF) that accounts for large-scale fading effects and imperfect power control. The CDF was generated following the channel modeling guidelines specified in~\cite{3GPP}. We repeat the same setup as Fig.~\ref{fig_1}, but with user streams encoded by a rate-1/2 convolutional code. In this simulation, perfect CSI is assumed at the base station. In plotting the coded BER versus average $E_{\text{b}}/N_{0}$ in Fig.~\ref{fig_3}, an individual detection and decoding scheme is adopted. The decoding is performed via soft Viterbi algorithm.
Looking at the coded BER in Fig.~\ref{fig_3}, significant gain of about 1 dB for M-BLAST (with $t=5$ iterations) over the linear MMSE is observed.
Typically, iterative schemes like the conventional PIC are known to be sensitive to large-scale fading. This is illustrated in Fig.~\ref{fig_3} by the inferior performance of the soft SIC, lagging behind that of the linear MMSE. Moreover, the improved robustness of the iterative M-BLAST to large-scale fading is evident in Fig.~\ref{fig_3}. This improved robustness may be attributed to the positive effect of the additional iterative equation~\eqref{eq_o}. \comment{Similar relative gain of M-BLAST over MMSE was also observed for the configuration of $K=32$ users and $M=96$ receiving antennas (about 0.5 dB gain) and when repeating the simulation under imperfect CSI and non-ideal noise variance estimation.}

Next, the achievable uplink user throughput is evaluated for the various detectors in the following manner. First, the simulated (with $K=500$, $M=1000$) post-detection signal-to-interference-and-noise-ratios (SINR) for the different detectors are plugged into Shannon's SISO-AWGN capacity equation, serving for our purposes as an upper bound on the user throughput.
Second, the CDF of a user's uplink SINR in a multi-cell network is obtained from two sources: 1) a multi-cell LTE 3D-UMi (urban micro-cell) channel~\cite{3GPP};
2) a massive MIMO uplink system (\cite{adhikary2017uplink}, and specifically Section~VII and Fig. 4 therein).
Now, we compare the $10$th (cell-edge), $50$th (median) and $90$th (center-cell) percentiles of the throughput distributions obtained with the various detectors. The relative increase in the throughput gained by M-BLAST w.r.t.~the legacy detectors is reported in Table~\ref{tab_1} under perfect (in \textbf{bold}) and imperfect CSI (non-bold) conditions. In addition to the MMSE-SIC, we have also evaluated here the ZF-based counterpart of V-BLAST.
Looking first at the LTE uplink use-case under ideal CSI conditions (upper row, in \textbf{bold}), M-BLAST can be seen to yield significant throughput gains, and thus may be beneficial for all the users in the system.
Under the realistic CSI conditions (non-bold), the trend in gains remains. For the median ($50\%$) throughput, although no gain is observed, M-BLAST still exhibits attractive computational benefit compared to MMSE and the impractical V-BLAST. It is interesting to note that M-BLAST gain versus the conventional soft PIC is larger in this case, pointing to its improved robustness under such pragmatic CSI conditions. In the massive MIMO use-case (lower row on Table~\ref{tab_1}), M-BLAST again exhibits similarly attractive behavior. Further relative gains for M-BLAST are observed under the imperfect CSI case. Note that the median SINR for the massive MIMO use-case is higher than its 3GPP-based equivalent, primarily because more antennas allow better suppression of the interference. Hence, the relative gains of M-BLAST over the legacy detectors (except PIC) in this case are more apparent. On the other hand, the advantage of M-BLAST over PIC in the intermediate, rather than high, SNR range (as shown in Fig.~\ref{fig_1}) leads to greater relative gains in the 3GPP use-case.

\begin{table}[!tb]
    \renewcommand{\arraystretch}{1}
\caption{\scriptsize{M-BLAST gains in throughput for perfect (in \textbf{bold}) and imperfect CSI.}}\vspace{-0cm}
\centering
\resizebox{0.55\columnwidth}{!}{%
\huge
\begin{tabular}{|c|c||c|c|c|c|c|c|c|c|}
  \hline
  \makecell{User's\\SINR CDF} & \% of users  & \multicolumn{2}{c|}{MMSE} & \multicolumn{2}{c|}{ZF-SIC} & \multicolumn{2}{c|}{MMSE-SIC} & \multicolumn{2}{c|}{PIC} \\ \hline
  \multirow{3}{*}{3GPP~\cite{3GPP}} &
  90\% & \textbf{24\%} & 20\% & \textbf{18\%}&15\% & \textbf{14\%}& 11\% & \textbf{2\%}& 10\% \\
  & 50\%  &\textbf{10\%}& 0\% & \textbf{13\%}& 5\% & \textbf{7\%}& 0\% & \textbf{10\%}& 18\% \\
  & 10\%  &\textbf{3\%}& 0\% & \textbf{25\%}& 25\% & \textbf{6\%}& 7\% & \textbf{9\%}& 32\% \\
  \hline
  \multirow{3}{*}{\makecell{Massive\\MIMO~\cite{adhikary2017uplink}}} &
  90\% & \textbf{21\%}& 25\% & \textbf{12\%}& 13\% & \textbf{10\%}& 12\% & \textbf{0\%}& 5\% \\
  & 50\%  & \textbf{27\%}& 26\% & \textbf{16\%}& 17\% & \textbf{13\%}& 14\% & \textbf{0\%}& 3\% \\
  & 10\%  & \textbf{1\%}& 0\% & \textbf{23\%}& 25\% & \textbf{6\%}& 8\% & \textbf{2\%}& 13\% \\
  \hline
\end{tabular}\label{tab_1}
}
\end{table}

\vspace{-0cm}
\section{Conclusion}\label{sec_conclusion}
This letter proposes an improved interference cancellation scheme, which is based upon parallel rather than successive detection architecture (as in the legacy V-BLAST). M-BLAST exhibits advantageous error performance along with computational efficiency, positioning itself as an attractive detection solution for large-scale MIMO applications. Note that in addition to the massive MIMO realm, M-BLAST may also be extremely beneficial for the Internet of Things (IoT) uplink, wherein a lot of users (devices) transmit simultaneously.\comment{The derived algorithm also calls for a new look at PIC-like design as an alternative to SIC, which is widely regarded as the detector of choice for non-orthogonal multiple-access (NOMA) on the uplink.} A study of M-BLAST architecture for higher constellations is currently underway.


\vspace{-0cm}

\begin{thebibliography}{10}
\providecommand{\url}[1]{#1}
\csname url@samestyle\endcsname
\providecommand{\newblock}{\relax}
\providecommand{\bibinfo}[2]{#2}
\providecommand{\BIBentrySTDinterwordspacing}{\spaceskip=0pt\relax}
\providecommand{\BIBentryALTinterwordstretchfactor}{4}
\providecommand{\BIBentryALTinterwordspacing}{\spaceskip=\fontdimen2\font plus
\BIBentryALTinterwordstretchfactor\fontdimen3\font minus
  \fontdimen4\font\relax}
\providecommand{\BIBforeignlanguage}[2]{{%
\expandafter\ifx\csname l@#1\endcsname\relax
\typeout{** WARNING: IEEEtran.bst: No hyphenation pattern has been}%
\typeout{** loaded for the language `#1'. Using the pattern for}%
\typeout{** the default language instead.}%
\else
\language=\csname l@#1\endcsname
\fi
#2}}
\providecommand{\BIBdecl}{\relax}
\BIBdecl

\bibitem{wolniansky1998v}
P.~W. Wolniansky, G.~J. Foschini, G.~Golden, and R.~A. Valenzuela, ``{V-BLAST}:
  An architecture for realizing very high data rates over the rich-scattering
  wireless channel,'' in \emph{Signals, Systems, and Electronics, 1998. ISSSE
  98. 1998 URSI International Symposium on}.\hskip 1em plus 0.5em minus
  0.4em\relax IEEE, 1998, pp. 295--300.

\bibitem{marzetta2010noncooperative}
T.~L. Marzetta, ``Noncooperative cellular wireless with unlimited numbers of
  base station antennas,'' \emph{{IEEE} Trans. Wireless Commun.}, vol.~9,
  no.~11, pp. 3590--3600, 2010.

\bibitem{marzetta2016fundamentals}
T.~L. Marzetta, E.~G. Larsson, H.~Yang, and H.~Q. Ngo, \emph{Fundamentals of
  Massive MIMO}.\hskip 1em plus 0.5em minus 0.4em\relax Cambridge University
  Press, 2016.

\bibitem{varanasi1990multistage}
M.~K. Varanasi and B.~Aazhang, ``Multistage detection in asynchronous
  code-division multiple-access communications,'' \emph{{IEEE} Trans. Commun.},
  vol.~38, no.~4, pp. 509--519, 1990.

\bibitem{tanaka2005approximate}
T.~Tanaka and M.~Okada, ``Approximate belief propagation, density evolution,
  and statistical neurodynamics for {CDMA} multiuser detection,'' \emph{{IEEE}
  Trans. Inf. Theory}, vol.~51, no.~2, pp. 700--706, 2005.

\bibitem{yedidia2001idiosyncratic}
J.~Yedidia, ``An idiosyncratic journey beyond mean field theory,''
  \emph{Advanced mean field methods: Theory and practice}, pp. 21--36, 2001.

\bibitem{plefka1982convergence}
T.~Plefka, ``Convergence condition of the {TAP} equation for the
  infinite-ranged {Ising} spin glass model,'' \emph{Journal of Physics A:
  Mathematical and general}, vol.~15, no.~6, p. 1971, 1982.

\bibitem{som2010improved}
P.~Som, T.~Datta, A.~Chockalingam, and B.~S. Rajan, ``Improved large-{MIMO}
  detection based on damped belief propagation,'' in \emph{Information Theory
  (ITW 2010, Cairo), 2010 IEEE Information Theory Workshop on}.\hskip 1em plus
  0.5em minus 0.4em\relax IEEE, 2010, pp. 1--5.

\bibitem{3GPP}
``Study on {3D} channel model for {LTE},'' 3GPP, Tech. Rep. 36.873, 2017,
  v12.6.0.

\bibitem{adhikary2017uplink}
A.~Adhikary, A.~Ashikhmin, and T.~L. Marzetta, ``Uplink interference reduction
  in large-scale antenna systems,'' \emph{{IEEE} Trans. Commun.}, vol.~65,
  no.~5, pp. 2194--2206, 2017.

\end{thebibliography}
\end{document}